# Strong Vibrational Relaxation of NO Scattered from Au(111): Importance of an Accurate Adiabatic Potential Energy Surface

Rongrong Yin, Yaolong Zhang, and Bin Jiang*

*Hefei National Laboratory for Physical Science at the Microscale, Department of Chemical Physics, Key Laboratory of Surface and Interface Chemistry and Energy Catalysis of Anhui Higher Education Institutes, University of Science and Technology of China, Hefei, Anhui 230026, China*

Experimental observations of multi-quantum relaxation of highly vibrationally excited NO scattering from Au(111) are a benchmark for the breakdown of Born-Oppenheimer approximation in molecule-surface systems. This remarkable vibrational inelasticity was long thought to be almost exclusively mediated by electron transfer; but, no theories have quantitatively reproduced various experimental data. This was suggested to be due to errors in the adiabatic potential energy surface (PES) used in those studies. Here, we investigate electronically adiabatic molecular dynamics of this system with a globally accurate high dimensional PES, newly developed with neural networks from first principles. The NO vibrational energy loss is much larger than that on earlier adiabatic PES. Additionally, the translational inelasticity and translational energy dependence of vibrational inelasticity are also more accurately reproduced. There is reason to be optimistic that electronically nonadiabatic theories using this adiabatic PES as a starting point might accurately reproduce experimental results on this important system.

Achieving a predictive understanding of heterogeneous catalysis is an ultimate goal of surface chemists and a grand challenge of physical chemistry.[1] This has motivated a variety of molecular beam experiments to investigate very fundamental aspects on the interactions and energy transfer between molecules and solid surfaces, at the quantum-state level.[2-5] These experimental data have been used to promote the development of theory and assess the validity of theoretical models,[6-10] which in turn enable a detailed mechanistic understanding of experimental findings from first principles. This interplay between these elegant experimental and theoretical works has revealed the reaction mechanisms in great detail for several prototypical molecule-surface systems.[11]

In this aspect, inelastic scattering of NO on Au(111) represents one of the most extensively studied benchmark systems that illustrates the breakdown of Born-Oppenheimer (BO) approximation (BOA) at metal surfaces.[12-14] Pioneering experiments by Wodtke and coworkers have discovered surprising multiquantum vibrational relaxation during the scattering of highly excited NO molecule from Au(111),[15-16] providing unambiguous evidence for vibrational promotion of electron transfer. This interesting non-adiabatic phenomenon has later inspired several different theoretical models,[17-21] which have semi-quantitatively reproduced the observed dramatic vibrational relaxation from NO($v_i$=15) scattering at a low translational energy of incidence ($E_i$) of ~0.05 eV.[15] More recent state-to-state measurements have supplied abundant vibrational excitation/deexcitation data of NO scattering from Au(111) with various initial states in a wide range of translational energies, providing more stringent tests for theoretical models.[14, 22-27] Nevertheless, none of existing theoretical models were able to reproduce the translational energy dependence of vibrational relaxation probabilities of NO($v_i$=3)[25] and vibrational state distributions of NO($v_i$=11) and NO($v_i$=16) at high translational energies.[27] It was argued that the disagreement is due largely to the insufficient accuracy of the adiabatic potential energy surface (PES), even though the dynamics of this system is dominantly non-adiabatic.[24-25, 27]

Indeed, all previous theoretical models relied on approximate PESs only. For example, the earliest Monte-Carlo stochastic wave packet model developed by Lin and Guo enabled quantum jumps between empirical neutral and negative ion states of the NO molecule.[17] Applying an electronic friction (EF) model, Monturet and Saalfrank constructed a two-dimensional ground state PES based on density functional theory (DFT) calculations and performed fully quantum-mechanical calculations based on an open-system density-matrix theory.[21] Tully and coworkers advanced an independent electron surface hopping (IESH) model for this system,[18] with a DFT based Newns-Anderson Hamiltonian.[20] However, only a few hundred ground state DFT points were computed for parametrizing the diabatic Hamiltonian expressed by simple pairwise potentials and the lattice motion was simply described by a generalized harmonic potential.[20] Although IESH calculations have successfully reproduced some of experimental findings,[19, 22] as mentioned above, the adiabatic PES obtained by diagonalizing the diabatic Hamiltonian was suggested to be "too-soft" and "too-corrugated",[25] without covering the NO dissociation channel.[25-26] A more accurate adiabatic PES built from first principles is therefore highly desirable.

To meet this challenge, in this Letter, we report the first globally accurate high-dimensional adiabatic PES that describes both NO scattering and dissociation on a mobile Au(111) surface, based on a faithful neural network (NN) representation of thousands of DFT energy and force data. We took our recently developed strategy[28] that combines a direct dynamics sampling procedure and an atomistic neural networks (AtNN) approach.[29] The AtNN representation expands the total energy of the system as the sum of atom dependent NNs described by their chemical environments. For each atom, the chemical environment is represented by the atom centered and element dependent two-body and three-body symmetry functions. Since it scales linearly as the number of atoms in the system, this powerful AtNN approach has been successfully used by us and others to develop molecule-surface PESs involving both molecular and surface degrees of freedom (DOFs).[28, 30-32] Direct dynamics trajectories were carried out on-the-fly at the spin-polarized DFT level using the Vienna Ab initio Simulation Package,[33-34] using the PW91 functional within the generalized gradient approximation (GGA).[35] Core electron interactions were described with the projector augmented wave (PAW) method.[36] The Au(111) surface was represented by a four-layer slab model in a 3×3 unit cell with the top two layers movable. The Brillouin zone was sampled by a 4×4×1 Gamma-centered $k$-point grid and the kinetic energy cutoff of the plane-wave basis was 400 eV. Methfessel-Paxton smearing method[37] for finite temperature calculations was applied with a smearing width of 0.2 eV. A dipole correction to the total energy was imposed along the lattice vector in Z direction to diminish the interactions between the vertically repeated images. These trajectories mostly explored the configuration space for molecular scattering but with no access to the transition state and dissociation channel, owing to the very large and tight dissociation barrier. We thus added dozens of points along the minimum energy path and randomly sampled hundreds of points in the dissociation channel. The final AtNN PES was trained to fit 2722 points with both energies and forces, in which the root mean square errors (RMSEs) are 33.3 meV for energies (1.67 meV/atom) and 31.6 meV/Å for forces, respectively. The AtNN PES reproduces the geometries and energies of stationary points quite well (see Figure S2 and Table S1). More computational details and more comprehensive accuracy checks can be found in the Supporting Information (SI).

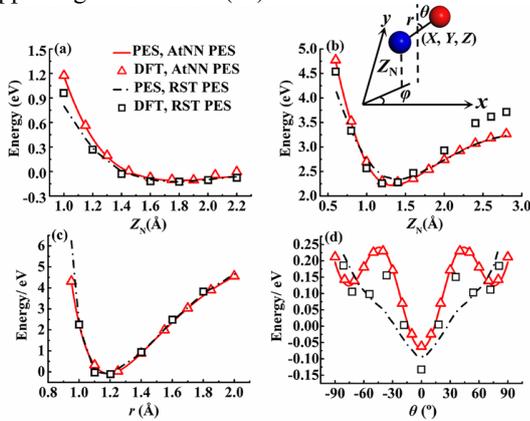

Figure 1. Comparison of AtNN (red solid curves) and RST PESs (black dash-doted curves) of the NO +Au(111) system, along with the corresponding DFT data in this work (red triangles) and Ref. [20] (black squares), as a function of $Z_N$ with $r$=1.191Å (a) and $r$=1.60 Å (b), a function of $r$ with $Z_N$ fixed at 1.60 Å (c), and a function of $\theta$ with $Z$=2.2Å and $r$=1.191 Å (d), respectively. We defined N-O bond length ($r$), the distance of N-surface ($Z_N$), the distance of the molecular center of mass to surface ($Z$), the polar angle ($\theta$) in an inset of panel (b). The NO molecule is placed on top of a hcp site at the frozen surface. Color scheme: blue = N, red = O. Note that the original data extracted from Ref. [20] may contain some minor digitization errors, which should not affect the comparison in a significant way.

Figure 1 compares several representative one-dimensional potential energy curves taken from the AtNN PES and the adiabatic Roy-Shenvi-Tully (RST) PES reported in Ref. [20], along with the original DFT points. This comparison is two-fold regarding the interaction between NO and the static surface. First, the DFT energies in the two studies are somewhat different. One possible reason is that the (3×3) supercell here is larger than the c(2×4) supercell in Ref. [20], corresponding to the NO coverage of 1/9 and 1/4 ML, respectively. In addition, a 4×4×1 Gamma-centered $k$-point grid is used in this work, compared to a 4×3×1 Monkhorst–Pack $k$-point grid in Ref. [20]. Specifically, from Figures 1(a)-1(b), it is found that the current DFT energy curves are slightly more repulsive as the NO molecule approaches the surface so that the AtNN PES is likely more favorable for translational inelasticity. Second, the AtNN PES reproduces original DFT energies much more accurately than the RST PES. This is not surprising given the very limited parameters and flexibility of the pairwise RST PES.[20] Indeed, the average prediction error of the RST PES with respect to 200 randomly selected configurations was ~120 meV,[18] which is much larger than that of the AtNN PES with respect to thousands of points. In particular, it is seen in Figure 1(d) that the RST PES predicts a simple orientational preference for NO being perpendicular to the surface, while the AtNN PES perfectly captures the subtle anisotropy in a very small energy range.

A more explicit comparison on the energy transfer and product energy disposal on the two PESs is enabled by quasi-classical trajectory (QCT) calculations, where the vibrational action number $v_f$ was determined by Einstein−Brillouin−Keller (EBK) semi-classical quantization and rotational quantum number $j_f$ by the quantum mechanical expression for rotational angular momentum. To partially account for the electronic excitation, we took advantage of the electronic friction (EF) model,[38] which effectively takes non-adiabatic effects due to the continuous excited states at metal surfaces into account, in a generalized Langevin equation on the adiabatic PES. The friction coefficient for each atom was determined within the local density friction approximation (LDFA),[39] which is dependent on the embedded electron density at the atomic position on the bare surface.[39] To make this evaluation possible, the electron density surface as a function of surface structure has also been analytically represented by AtNN. More technical details of our QCT simulations can be found in the SI.

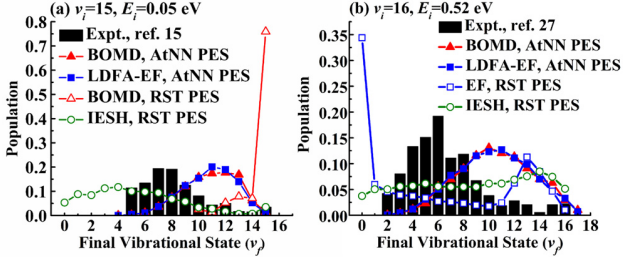

Figure 2. Comparison of experimental vibrational state distributions[15, 27] of NO scattering from Au(111) (black bar), with BOMD (red filled triangle) and LDFA-based EF ones in this work (blue filled square), BOMD[19] (red open triangle), EF[27] (blue, open square), and IESH[19, 27] (green open circle) ones using the RST PES, for (a) NO($v_i$=15), $E_i$=0.05 eV and (b) NO($v_i$=16), $E_i$=0.52 eV, with the surface temperature ($T_S$) of 300 K.

In Figures 2(a)-2(b), the final vibrational state ($v_f$) distributions of highly excited NO molecules scattering from Au(111) calculated on the AtNN PES are compared with previous theoretical[19, 27] and experimental[15, 27] data. At $v_i$=15 and $E_i$=0.05 eV, Born Oppenheimer molecular dynamics (BOMD) results on the AtNN PES reveal significant vibrational energy transferred adiabatically to other DOFs, in sharp contrast with the little adiabatic vibrational energy loss predicted previously.[19] More specifically, in Figure S3, it is found that ~0.46 eV vibrational energy is dissipated to the lattice and there is also some intramolecular energy transfer from vibration to translation (~0.20 eV) and rotation (~0.14 eV). This renders an overall ~0.8 eV vibrational energy loss and a very broad final vibrational state distribution peaking at $v_f$=11~12. This amounts to roughly half of experimental observation of the deexcitation of 7~8 vibrational quantum numbers and ~1.5 eV energy loss.[15] These results imply that the semi-quantitative agreement between IESH and experimental results under this condition could be a coincidence, as has been noted by Krüger et al.[26] Comparison is also shown in Figure 2(b) for NO($v_i$=16) at $E_i$=0.52 eV, where neither the IESH nor EF model on the RST PES reproduces the experimental data.[27] Note that BOMD results were not reported in Ref. [27], but they should be similar to the EF ones (e.g., see Ref. [22]), which feature an abrupt peak at $v_f$=0, in qualitative contradiction with the experiment. However, it should be noted that scattered NO($v$=0) can not be reliably detected in the experiments as the signal is overlapped by the incident NO($v$=0) molecules, so experimental population of the ground state remains unclear. On the other side, BOMD results on the AtNN PES again show a substantial decrease of vibrational energy, populating many lower vibrational states with a peak at $v_f$=10. The comparison of AtNN results with experimental distributions in Figure 2 shows that with an accurate adiabatic PES, large amounts of vibrational relaxation occur within an adiabatic dynamics model. Therefore, one would expect that an accurate treatment of electronic nonadiabaticity with this PES might better agree with experiment. We note that the DFT-GGA predicted adsorption energies of NO on several transition metal surfaces (not including Au(111) though) suffer from systematic errors due to an overestimation of back-donation charge transfer[40]. Given the importance of the adiabatic PES in describing vibrational relaxation, further improvement could be expected if DFT is replaced by more advanced electronic structure methods such as quantum mechanical embedding theory[41] or quantum Monte Carlo approach[42]. The nuclear quantum effects neglected in our QCT model should also be examined with more advanced quantum dynamics methods. Unfortunately, they are both too expensive at present. It should be noted that the electronic nonadiabaticity in this system is hardly captured by the LDFA-based EF model, which results in minor nonadiabatic energy transfer and yields very similar results as adiabatic ones, in accord with previous theoretical findings.[22, 27] We will therefore mainly discuss our adiabatic results below.

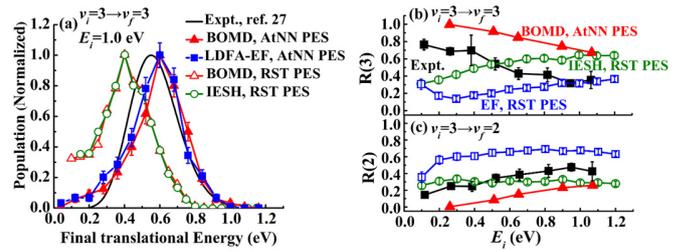

Figure 3 (a) Comparison of experimental final translational energy distributions[27] (black curve) of NO($v_i$=3 →$v_f$=3) scattering from Au(111) at $E_i$=1.0eV and $T_S$=320 K, with theoretical results from BOMD (red filled triangle) and LDFA-EF (blue filled squares) in this work, and BOMD (red open triangle) and IESH (green open circle) simulations in Ref. [27]. (b-c) Experimental branching ratios (black filled square) of (b) of NO($v_f$=3) and (c) NO($v_f$=2) scattered from NO($v_i$=3) are compared with BOMD and EF ones in this work, and EF (blue open square) and IESH (green open circle) results in Ref. [25], as a function of $E_i$ at $T_S$=300 K. The branching ratio is defined R($v_f$)=S($v_f$)/(S(1)+S(2)+S(3)), where the S($v_f$) is the scattering probability to a final vibrational state ($v_f$).

Next, we compare the final translational energy distribution in the vibrational elastic channel for ($v_i$=3→$v_f$=3), as a check of translational inelasticity. As shown in Figure 3(a), both previous BOMD and IESH calculations on the RST PES predict similarly too low translational energies of the scattered NO molecules, shifting the translational energy distributions to the low energy end by ~0.2 eV. The RST PES was thus argued to be "too-soft" for too facile energy transfer from translation to other DOFs during scattering.[25] In contrast, the adiabatic AtNN PES reproduces the measured translational energy distribution[27] fairly well. In Figure 3(b), the measured branching ratios of NO($v_f$=3) and NO($v_f$=2) after scattering from NO($v_i$=3)[25] are compared with different calculated results as a function of translational energy. Clearly, both IESH and EF calculations using the RST PES present the opposite dependence of vibrational relaxation of NO($v_i$=3) on the translational energy. On the contrary, the BOMD results on the AtNN PES reproduce well the increasing vibrational inelasticity with translational energy, although they underestimate the absolute vibrational deexcitation probabilities. The adiabatic dynamics results suggest that the direct mechanical vibrational energy loss to phonons increases with the translational energy, in accord with the experimental trend. Further nonadiabatic dynamics calculations based on this PES might better agree with experiment.

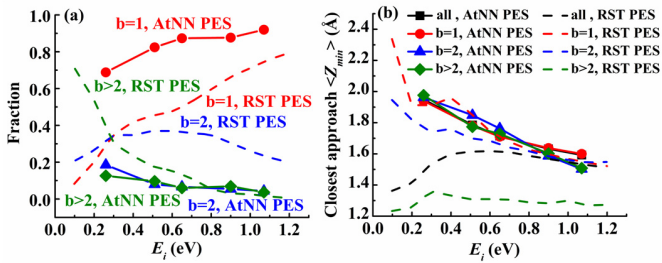

Figure 4 (a) Fraction of trajectories with single bounce (b=1), double bounce (b=2) and multiple bounce (b>2), as a function of $E_i$ for adiabatic simulations in this work (solid lines with symbols) and in Ref. [25] (dashed lines), with NO($v_i$=3) scattering from Au(111) at $T_s$=300 K. (b) The average closest vertical distance $<Z_{min}>$ of NO above the top metal layer, evaluated from single bounce (red, filled circle), double bounce (blue, filled triangle), multiple bounce (green, filled diamonds), and all trajectories (black, filled square) in this work and in Ref. [25].

The aforementioned results unambiguously demonstrate that an accurate adiabatic PES is a prerequisite in order to correctly describe this benchmark non-adiabatic system. To gain a deeper insight, it is interesting to explore why the adiabatic AtNN PES outperforms the RST counterpart. Interestingly, Golibrzuch et al. found that the RST PES features too many bounces of NO on the surface which enhance the non-adiabatic energy transfer.[25] The high fraction of multi-bounce collisions results from the artificial attractiveness of the RST PES. To check this feature on the more realistic AtNN PES, we count a bounce in a scattering trajectory whenever the center-of-mass velocity of NO along the z axis reverses its sign. Our results extracted from about 1000 representative BOMD trajectories are compared to those reported by Golibrzuch et al. in Figure 4(a). The only similarity of the two results is that the fraction of single bounce always increases with $E_i$, consistent with the direct scattering mechanism. Remarkably, the fraction of single bounce trajectories on the AtNN PES is dominant regardless the translational energy, much higher than that on the adiabatic RST PES. This suggests that the AtNN PES is less attractive enabling more direct scattering trajectories. In addition, Figure 4b compares closest distance of NO above the surface, namely $<Z_{min}>$, averaged over numerous trajectories on the two PESs as a function of $E_i$. Clearly, the NO molecule approaches closer to the surface on the RST PES, especially at low energies. Moreover, because of the multi-bounce contribution, $<Z_{min}>$ decreases with the decreasing $E_i$ on the RST PES at low energies. This is an indication of a too strong steering effect, as typically a molecule with a smaller translational energy is likely to be more easily reflected in the sudden limit. This ability of accessing to the high electron density area would enlarge the non-adiabatic vibrational-to-electronic energy transfer and overcompensate the underestimated adiabatic vibrational energy loss on the RST PES.[25]

Another possible defect affecting the dynamics is the absence of NO dissociation channel in the RST PES, as argued by Schäfer and coworkers, because the highly vibrationally excited NO molecule carries sufficient energy for dissociation.[27, 43] As shown in Figure S2, the AtNN PES captures the NO dissociation channel and predicts a dissociation barrier of 2.88 eV, which is actually much lower than the available vibrational energy of, e.g., NO($v_i$=16) plus a translational energy. Nevertheless, no reactive trajectory was found in any calculations performed in this work. This extremely low reactivity can be ascribed to the entropic effects stemming from the too tight transition state,[11, 44] which morphs the shape of the PES from the entrance channel to the transition state and favors a bottleneck for dissociation. This topography is radically different from that of a completely non-reactive PES, e.g. for NO approaching the surface perpendicularly (see Figure S5). As a result, the adiabatic vibrational energy transfer is promoted as NO climbs up to the transition state but fails to dissociate.[27] The influence of the reactive part of the PES at energies below the saddle point on the vibrational energy transfer has been already discussed by Rettner et al., who have found that the direct vibrational excitation of $H_2$ on Cu(111) is governed by the same region of the PES as dissociative adsorption.[45] Despite the fact that our results do not show any dissociative adsorption of NO, they are critically important to eventually be able to describe the influence of BOA failure on surface reaction probabilities, a subject of future work. For example, future studies on this system at higher total energies are likely to lead to dissociation. Similar systems like NO on Ag and NO on Cu,[46-47] where the methods presented here can also be applied, are likely to be more reactive at energies that are experimentally accessible.

We close the discussion by addressing the influence of surface reconstruction. The Au(111) surface is well known to form a 22×$\sqrt{3}$ "herringbone" reconstruction,[48] with 23 surface atoms fit into 22 lattice sites by compressing the top layer of the surface. Describing strictly this reconstructed structure requires a very large supercell that is too expensive for DFT and dynamics calculations. Fortunately, a recent DFT study suggests that the neglect of local reconstruction of Au(111) typically involves a small change of gas-surface interaction energy (well below 100 meV for a single chemisorption bond).[49] This difference is unlikely to result in any qualitative physical changes in most cases. In addition, Janke et al. found almost the same energy loss profile of H atom scattering on Au(111) using either a flat Au(111) or a "herringbone" reconstructed Au(111) supercell in simulations[50], Wijzenbroek et al. also found that the difference between the dissociation barriers for $H_2$ on the reconstructed and unreconstructed Au(111) surfaces is within 26~90 meV[51]. These energy differences are unlikely to result in any qualitative physical changes in most cases. As a result, the use of unreconstructed Au(111) in this work is reasonably justified and should not alter our main conclusion in a qualitative way.

To summarize, we report a globally accurate full-dimensional AtNN PES based on a large number of DFT calculations that describes both NO scattering and dissociation on a mobile Au(111) surface. Its high efficiency (~$10^4$ times faster than DFT) enables extensive BOMD simulations, which lead to much more significant adiabatic vibrational energy loss than previous theoretical models. The better performance of the AtNN PES can be attributed to the more reliable description of the repulsive molecule-surface interaction and the change of potential energy topography due to the tight transition state, enabling more facile adiabatic vibrational energy transfer and more favorable translational inelasticity. Our results underscore the importance of the accurate potential energy landscape in determining molecular scattering dynamics at a metal surface. It should be noted that DFT energies have some uncertainties (e.g. dissociation barrier heights depend on the density functionals used), which would however not change significantly the topography of PES. Therefore, dynamics results presented here are not expected to be altered qualitatively. The LDFA based EF treatment with an AtNN-based embedded density surface only brings in a very minor non-adiabatic energy loss. The current PES serves as a good starting point towards developing more advanced first-principles non-adiabatic models of this and other relevant systems that might well describe experimental observations. With this approach, one can be optimistic that we will be able to reveal a clear understanding of the role of BOA failure in chemical reactions at metal surfaces.

**Acknowledgements**: We appreciate the support from National Key R&D Program of China (2017YFA0303500)National Natural Science Foundation of China (21722306, 91645202, and 21573203), Anhui Initiative in Quantum Information Technologies (AHY090200). We thank Maite Alducin and Hua Guo for some helpful discussion.